\documentclass[letterpaper, 10 pt, conference]{ieeeconf}  

\usepackage{hyperref}
\hypersetup{
    colorlinks=true,
    linkcolor=blue,
    filecolor=magenta,      
    urlcolor=cyan,
    pdftitle={Overleaf Example},
    pdfpagemode=FullScreen,
    }

\IEEEoverridecommandlockouts                              

\usepackage{graphics} 
\usepackage{epsfig} 
\usepackage{mathptmx} 
\usepackage{times} 
\usepackage{amsmath} 
\usepackage{amssymb}  
\usepackage[capitalise]{cleveref}
\usepackage{subcaption}

\usepackage{placeins}
\usepackage{lipsum}
\usepackage{makecell} 
\usepackage{tikz,pgfplots}
\usetikzlibrary{calc}

\newcommand\copyrighttext{%
	\footnotesize \textcopyright2018 IEEE:
	Personal use of this material is permitted. Permission from IEEE must be obtained for all other uses, in
	any current or future media, including reprinting/republishing this material for advertising or promotional
	purposes, creating new collective works, for resale or redistribution to servers or lists, or reuse of any
	copyrighted component of this work in other works.}
\newcommand\copyrightnotice{%
	\begin{tikzpicture}[remember picture,overlay]
		\node[anchor=south,yshift=10pt] at (current page.south) {\fbox{\parbox{\dimexpr\textwidth-\fboxsep-\fboxrule\relax}{\copyrighttext}}};
	\end{tikzpicture}%
}

\title{\LARGE \bf
Performance of Graph Database Management Systems as route planning solutions for different data and usage characteristics
}

\author{Karin Festl$^{1}$, Patrick Promitzer$^{1}$, Daniel Watzenig$^{1}$ and Huilin Yin$^{2}$
\thanks{This paper is part of the AI4CSM project that has received funding from the ECSEL Joint Undertaking (JU) under grant agreement
No 101007326. The JU receives support from the European Union’s Horizon 2020 research and innovation
programme and Germany, Austria, Belgium, Czech Republic, Italy, Netherlands, Lithuania, Latvia,
Norway”. In Austria the project was also funded by the program “IKT der Zukunft” of the Austrian Federal Ministry for Climate Action (BMK).
The publication was written at Virtual Vehicle Research GmbH in Graz and partially funded within the
COMET K2 Competence Centers for Excellent Technologies from the Austrian Federal Ministry for Climate Action (BMK),
the Austrian Federal Ministry for Labour and Economy (BMAW), the Province of Styria (Dept. 12)
and the Styrian Business Promotion Agency (SFG). The Austrian Research Promotion Agency (FFG) has been authorised for the programme management.}
\thanks{\authorrefmark{1}Virtual Vehicle Research GmbH, Graz, Austria {\tt\small karin.festl@v2c2.at}}%
\thanks{$^{2}$ School of Electronic and Information Engineering, Tongji University, Shanghai, China. }%
}

\begin{document}

\maketitle
\copyrightnotice
\thispagestyle{empty}
\pagestyle{empty}

\begin{abstract}
Graph databases have grown in popularity in recent years as they are able to efficiently store and query complex relationships between data. Incidentally, navigation data and road networks can be processed, sampled or modified efficiently when stored as a graph. As a result, graph databases are a solution for solving route planning tasks that comes more and more to the attention of developers of autonomous vehicles. To achieve a computational performance that enables the realization of route planning on large road networks or for a great number of agents concurrently, several aspects need to be considered in the design of the solution. Based on a concrete use case for centralized route planning, we discuss the characteristics and properties of a use case that can significantly influence the computational effort or efficiency of the database management system. Subsequently we evaluate the performance of both Neo4j® and ArangoDB depending on these properties. With these results, it is not only possible to choose the most suitable database system but also to improve the resulting performance by addressing relevant aspects in the design of the application.


\end{abstract}

\section{INTRODUCTION}
With the increasing amount of data that is availalbe and processed in different areas of applications, Graph Database Management systems (GDBMSs) have become more popular than ever. More and more database vendors developed GDBMSs, among others also Oracle Property Graph \cite{oracle}. Following this trend, also the utilization of GDBMSs in autonomous driving, for example for scene construction, route planning, traffic flow prediction, risk analysis and data visualization has become more significant. Especially route planning has become more crucial than ever due to the uprising deployment of assisted and autonomous driving. These developments have also promoted vehicle-to-everything communication, providing a huge amount of data most of which can be mapped geographically.
To store and process this data in the ego vehicle or also in a centralized way (as conducted e.g. in \cite{Zambrano-Martinez2019}), the usage of GDBMSs can be useful. With this, it is possible to efficiently store and access data that can be represented as a graph. It is clear that geographic data such as road networks are well suited for a graph representation. Moreover, as any Graph Database Managment System, GDBMSs enable the handling of queries in a server-like way while guaranteeing ACID compliance \cite{db_acid}. Thus, GDBMS is a suitable solution to process map information in a centralized way. Agents can access the GDBMS for adding map information such as traffic jams and roadworks and for finding their optimal routes.

When working with large amounts of data and possibly a large amount of agents, efficiency is crucial. The performance of GDBMSs has been studied and evaluated multiple times in the past \cite{Rusu2019,Angles2012,Wang2020a}. In \cite{Patras2021} and \cite{Miler2014} the performance of a GDBMS compared to a relational database management system has been compared for the specific use case of route planning. In \cite{Garcia2022} a concept for managing map and traffic data originating from sensors and V2X communication is presented.
In these studies, real map data was used for the evaluation. However, it is still not clear how the characteristics of the the data influence the computation time of route finding and other queries.

The contribution of this paper is to evaluate the performance of two different GDBMSs, namely Neo4j® and ArangoDB, for the application of route finding and map data adaption. Moreover, we evaluate how this performance changes for different characteristics of the map data (such as the degree of the graph and the cost distribution) and also for different usage patterns (such as consecutive or sequential query submission). With these results, the performance can not only be addressed by choosing the most suitable GDBMS but also in other aspects of developing the routing solution, such as the representation of the road network in a graph and assessing scalability to multiple agents.

We start our study with an example of a routing application in a real city to illustrate the intended application. In section~\ref{sec:methodology}, the methodology for generating datasets and queries for the evaluation is described. In section~\ref{sec:results}, the efficiency of the 2 evaluated GDBMSs for varying parameters (map data characteristics and usage patterns) is presented and discussed. Section~\ref{sec:conclusion} summarizes the results and presents conclusions.




\section{Example application}
\begin{figure}
\label{fig:amberg_path_map_small}
  \includegraphics[width=0.45\textwidth, trim={15cm 20cm 25cm 20cm},clip]{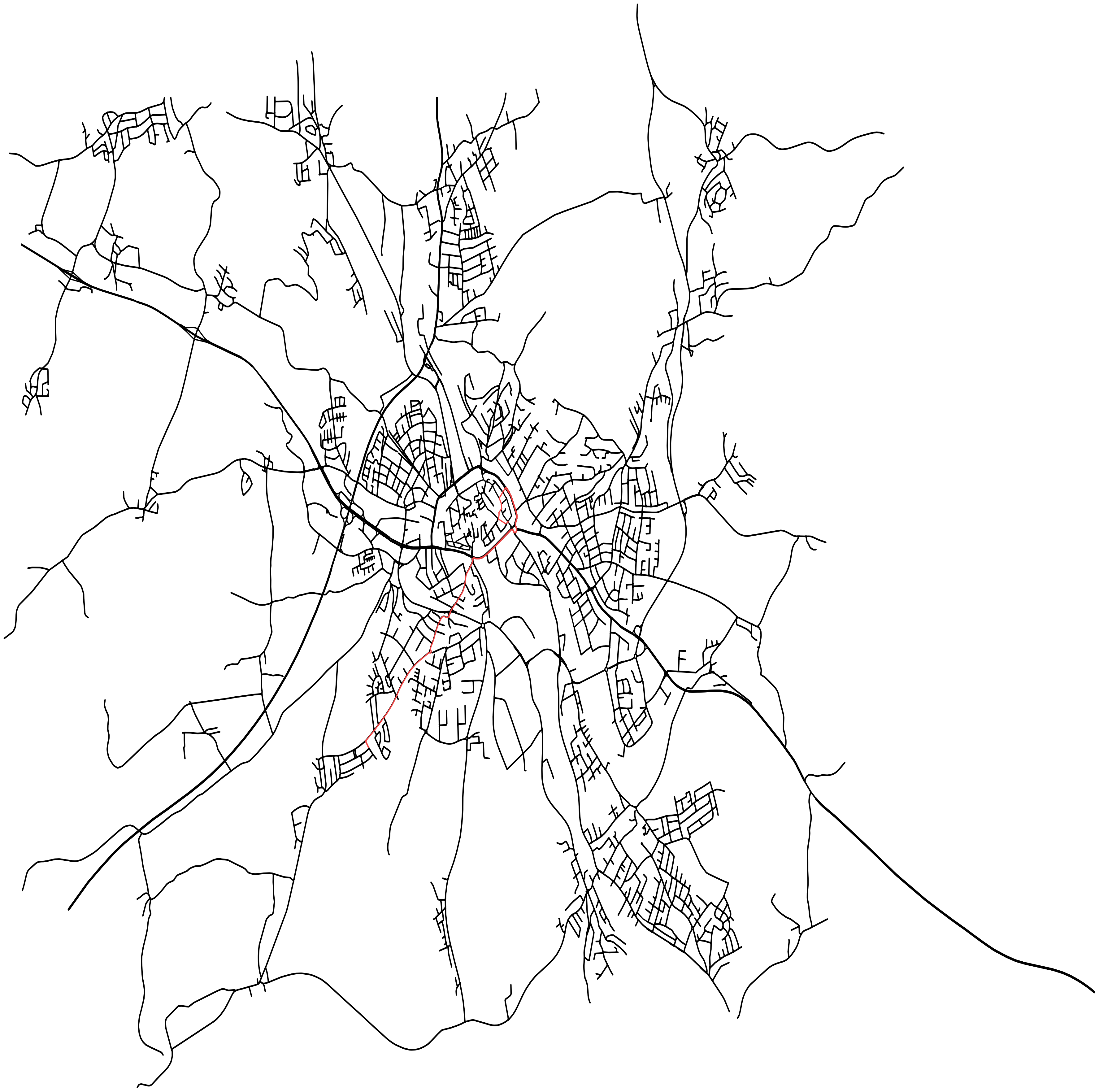}
  \caption{Road network of the city Amberg. In red is an exemplary shortest route from one position to another.}
  \label{fig:amberg_map}
\end{figure}

In our exemplary application, an autonomous taxi service shall be implemented in the city Amberg shown in Fig.~\ref{fig:amberg_map}. In order to make optimal use of the taxis (considering traveled distance, fuel consumption and time), a central unit is implemented to execute the following tasks:
\begin{itemize}
    \item assign customers to taxis
    \item plan routes for each taxi to their customer and to the customers goal positions
    \item update the map information to account for traffic jams and blocked roads
\end{itemize}

\begin{figure}[h]
	\centering
	\begin{subfigure}[b]{0.23\textwidth}
		\centering
		\includegraphics[width=\textwidth, trim=0 0.5cm 0 0,clip]{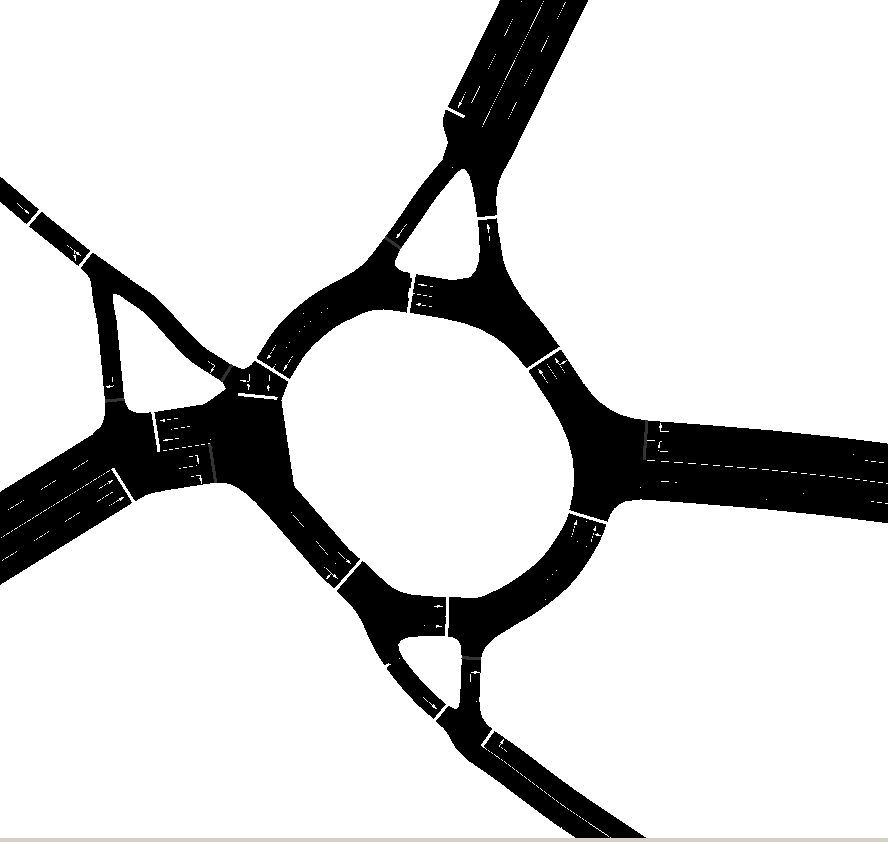}
		\caption{Opendrive visualization}
	\end{subfigure}
	\begin{subfigure}[b]{0.23\textwidth}
	\centering
		\includegraphics[width=\textwidth, trim=2cm 0 0 0,clip]{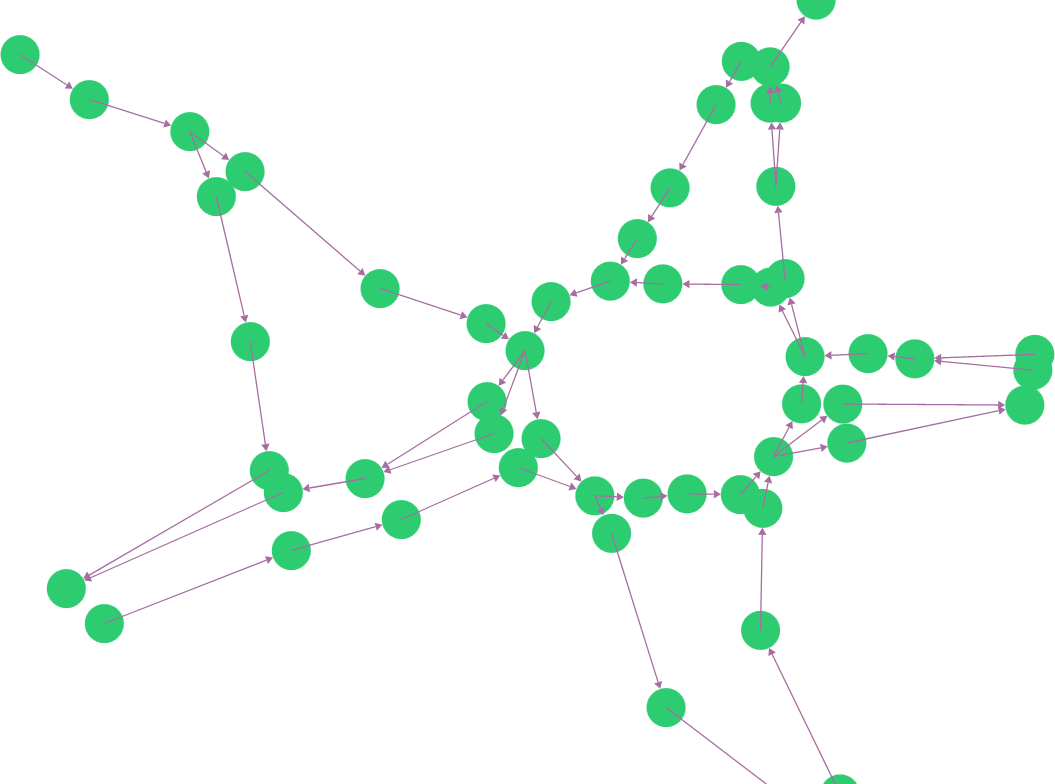}
	\caption{Graph representation}
	\end{subfigure}
\caption{An exemplary roundabout in the city Amberg.}
\label{fig:roundabout}
\end{figure}

To accomplish these tasks, a GDBMS is used. Thus, the first step is to convert the map information to a graph. The map information is gained from OpenStreetMap \cite{osm} and then converted to Opendrive \cite{odr} using the converter from Carla \cite{carla}.
We define for each Opendrive road element $r_i$ a node $n_i$ that is connected to all road elements that can be reached from $r_i$. The weight of a connection (or edge) from $n_i$ to another node $n_j$ represents the cost for traveling on the road segment $r_i$. Where the cost is a function of travel time, distance and fuel consumption. In Fig. \ref{fig:roundabout}, an exemplary roundabout is shown. With this representation, the map data of Amberg converts to a graph with $2.1\cdot10^4$ nodes and $3.0\cdot10^4$ edges.
It is typical for this kind of data, that a great portion of the nodes have only a single outgoing edge. While the graph could be easily reduced due to this property, it is favorable to keep the original structure. This way, dynamic changes in the map such as traffic jams can be implemented more realistically and the route can be directly translated back from the graph to the map. \par

The GDBMS queries required to acomplish the specified tasks are:
\begin{itemize}
    \item import the map data into the database
    \item find the shortest path between 2 given nodes (corresponding to the locations of the taxi and a costumer)
    \item modify the edges weights or delete them according to notifications from the taxis or roadside units (e.g. traffic jam or road blockade)
\end{itemize}

For importing the map data, both ArangoDB and Neo4j® provide the bulk import to efficiently load great amounts of data at once. For the presented map of Amberg, the bulk import takes approximately $2.7$s. To compute the shortest path, ArangoDB and Neo4j® both are using the Dijkstra algorithm. Querying a route in the presented map takes approximately $0.031$s for start- and goal positions that are $100$ nodes apart.

\section{Test methodology}\label{sec:methodology}
We want to evaluate the performance of the 2 GDBMSs for the queries that are used to implement the route planning application. These can be summarized to the loading of map data in form of a graph into the GDBMS and the finding of shortest paths in this graph.
The main goal is to quantify how the computation time changes depending on how the data is represented and how the queries are formulated. To eliminate coincidences in a specific test scene and to get representative results, the tests are repeated multiple times on different data sets.

In this section, we will first describe how the data is generated, then the usage patterns (in other words, how the queries are formulated) are explained and at the end of this section, test environment and procedure is explained.

\subsection{Generating the data sets}\label{sec:gendata}
\begin{figure}[h]
	\centering
	\begin{subfigure}[b]{0.23\textwidth}
		\centering
		\includegraphics[width=\textwidth, trim=0 0 0 0,clip]{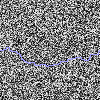}
		\caption{Uniform distribution}
	\end{subfigure}
	\begin{subfigure}[b]{0.23\textwidth}
	\centering
		\includegraphics[width=\textwidth, trim=0 0 0 0,clip]{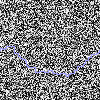}
	\caption{Beta distribution}
	\end{subfigure}
\caption{Mazes with different cost distribution. The optimal path (blue line) connects two positions with the minimum cost (the lightest colors in the illustration)}
\label{fig:cost_distribution}
\end{figure}

A data set for one test run consists of map data of variable size as well as a start and goal position for the path search.
To be able to vary the characteristics of the data set independently, the data is not retrieved from real maps but it is generated.
The map data is represented by a cell grid of square shape, as illustrated in Fig. \ref{fig:cost_distribution}, where the cells represent road segments of a map. For each grid cell we define a value $c_{i,j}\in [0,255]$ representing the cost for traversing this cell. Both the cost values and the start and goal position are generated randomly, where the euclidean distance between start and goal position is fixed.

In Fig. \ref{fig:cost_distribution}, exemplary mazes with two different cost distributions are shown. The beta distribution is parameterized such that most edges have very large or very low costs. This is a typical distribution in unstructured environments, where the grid cells do not represent road segments but geometric areas in the environment. These cells are then either occupied (large cost) or free (low cost). In the other tested variant, the cost is uniform distributed. This better resembles a road network, where sections can vary in length and traveling effort arbitrarily. Sections of very short length (for example where the road is split in small sections to enable detailed definitions of road properties), medium length (road sections with no special characteristics) and large effort (for example long sections on a motorway or sections with traffic jam or other obstacles) do occur with the same probability.

The size of the graph may increase due to a larger region that is considered or also due to a higher level of detail. The former adds both nodes and edges in a relation corresponding to the characteristic of the road network, which will not necessarily change the edges-per-node ratio. In the latter, the road segments are split in sections, creating additional road segments that are linked to exactly 2 road segments (nodes with exactly 2 edges). Hence, the edges-per-node ratio will draw closer to $2$. 
In the generated grid maze, all nodes (except for the border cells) have exactly $4$ neighbors to which they are connected bidirectional, resulting in $4$ edges per node. 
To vary the number of edges per node, we vary the dimension of the grid. The $1$D grid, with $2$ edges per node resembles a road that is sectioned to increase the level of detail. In the $3$D grid, each cell has $6$ neighbors resulting in $6$ edges per node. This data set can either resemble a road network with a high number of crossings, or a dynamic unstructured environment, where the third dimension represents time.


\subsection{Usage patterns}\label{sec:usage}
The GDBMS can be utilized for route planning in different ways, ranging from a simple navigation unit based on a static map to a centralized server that gathers traffic and road information from multiple sources to update the map in a way that optimal routes can be provided to all users. Depending on the specific application, the queries for importing data into the GDBMS and for solving the shortest path problem can differ in multiple ways.

\subsubsection{Concurrent queries}
Managing map data in a database offers many opportunities for cooperative and shared usage of this data. 
For example sharing information of the ego position and motion, it is possible to navigate efficiently within a multi-agent environment. One could also implement the possibility of sharing information about new roads, blocked roads or a change of traffic rules on parts of the roads.
When realizing these features, the GDBMS is required to handle multiple queries concurrently.\par
Both ArangoDB and Neo4j® are ACID compliant \cite{db_acid}, which guarantees that the concurrency of queries does not influence their results. However, it is to be evaluated, how well the database systems enable concurrency (i.e. if the queries are handled faster, when submitted concurrently). Therefore, we evaluate the processing time for multiple queries that are submitted sequentially in one test run and concurrently in another. To implement the concurrency, we generate $10$ threads that submit multiple queries each at the same time.\\
For importing data to the DB, the nodes and edges of one maze are divided in $10$ portions of equal size and distributed to the threads. For the path search $10$ start and end points are generated so each thread queries a different route.

\subsubsection{Cold and warm state}
When multiple queries are handled, the GDBMS increases its speed by caching data. At the beginning, when the cache is empty, the system is said to be in cold state. As more and more nodes are "touched" (they are accessed in order to handle a query), the cache fills and the system warms up. ArangoDB and also Neo4j® enterprise edition offer the possibility of an auto-warmup. In our evaluation, this option is deactivated.
To evaluate this difference, we perform tests in 4 different settings:
\begin{itemize}
    \item cold: the cache is cleared before submitting the query
    \item warm: the database is initialized before clearing the cache. Then, the a path-finding query is executed and only after this, the evaluation starts
    \item warmer: the setting is similar to warm, but 2 path-finding queries are executed before starting the evaluation
    \item hot: the database is initialized (all nodes and edges are added) and the evaluation starts immediately after (i.e. the path-finding query under test is submitted)
\end{itemize}
The cache and all RAM is cleared by restarting the docker container where the GDBMS runs in.

\subsubsection{Bulk and single import}
For importing the map data into the database, both GDBMSs provide the possibility of bulk data import. This is a more efficient way to add multiple nodes or edges instead of creating an extra query for each piece. However, when small pieces of data get available from multiple sources or over a period of time, the single data import queries are appropriate. Therefore, we evaluate different queries for importing data into the GDBMS.\\
\textbf{Single import}:
In Neo4j®, a single node is added to the DB with the query \texttt{CREATE}. With the query \texttt{MERGE} a single edge is added. In ArangoDB the query \texttt{INSERT} is used for both adding a single node and for adding a single edge.\\
\textbf{Bulk import}:
In Neo4j®, multiple nodes or multiple edges are compound to a bulk with \texttt{UNWIND}. Both bulks are then imported into the DB with \texttt{CREATE} or \texttt{MERGE}. In ArangoDB, the nodes and edges are compound to a list using \texttt{FOR} and then added to the graph using \texttt{INSERT}. Similar to Neo4j®, we use one query for importing the bulk of nodes and one for importing the bulk of edges.

\subsection{Testing environment and procedure}
The tests are executed on a computer running on Ubuntu Linux, release 20.04.5 LTS, a samsung SSD 980 PRO 2TB, an AMD Ryzen 7 2700x (8 cores) and $32$GB of RAM.
The GDBMS runs in a Docker container for which $24$GB of memory is reserved. Queries are generated outside the Docker container and sent over http/bolt.
The GDBMSs under test are
\begin{itemize}
    \item Neo4j® (neo4j:4.4.16-community)
    \item ArangoDB (arangodb:3.10.4)
\end{itemize}
\par

For quantifying the performance, we measure the execution time and the RAM usage. The time is measured from the point in time where the query is sent until the GDBMS responds with the result. The RAM usage is measured frequently during this time and its maximum value is accounted for evaluation.

To evaluate the performance of the GDBMS, tests are conducted multiple times with varying the parameters described in this section. 
One test run (i.e. the procedure to generate one data point in the results we are presenting) consists of the following procedure:
\begin{enumerate}
    \item generate or retrieve random data sets \label{enum:1}
    \item prepare the RAM as described in Sec. \ref{sec:usage} \label{enum:2}
    \item submit a query \label{enum:3}
    \item retrieve processing time and maximum RAM usage
    \item repeat \cref{enum:1,enum:2,enum:3} 10 times
    \item compute the average and standard deviation of processing time and RAM usage
\end{enumerate}
For each test run, $10$ different data sets are used. In all tests, the same data set is used for both GDBMSs. The standard deviation is visualized in the results only if the value exceeds $1\%$ of the average value.

For the shortest path queries, we use the Dijkstra algorithm with default parameters on both GDBMSs.

\section{Results}\label{sec:results}

To demonstrate the influence of each parameter on the performance, we conducted multiple series of tests, where one parameter is varied while the others are at the standard values defined in Tab.~\ref{tab:par_var}. 
We also evaluated the cross-correlation between all the parameters but can conclude, that there is no significant unexpected behavior. One exception is the degree of the graph, which changes the influence of the cache state. This is also described in more detail in this section. The cost distribution as described in Sec. \ref{sec:gendata} appears to have no influence on the performance and thus no dedicated results are presented.
\begin{table}[h]
\caption{Standard values used for the evaluation.}
\label{tab:par_var}
\begin{center}
\begin{tabular}{|l|l|}
\hline
Parameter & Value\\
\hline
Number of nodes & $10^4$\\
Edges per node & $4$\\
Cost distribution & beta\\
Concurrency & No\\
Cache & warm\\
Query type & bulk\\
\hline
\end{tabular}
\end{center}
\end{table}

\begin{figure}
	\centering
	\begin{subfigure}[b]{0.45\textwidth}
		\centering
		\includegraphics[width=\textwidth, trim=0 0cm 0 0cm,clip]{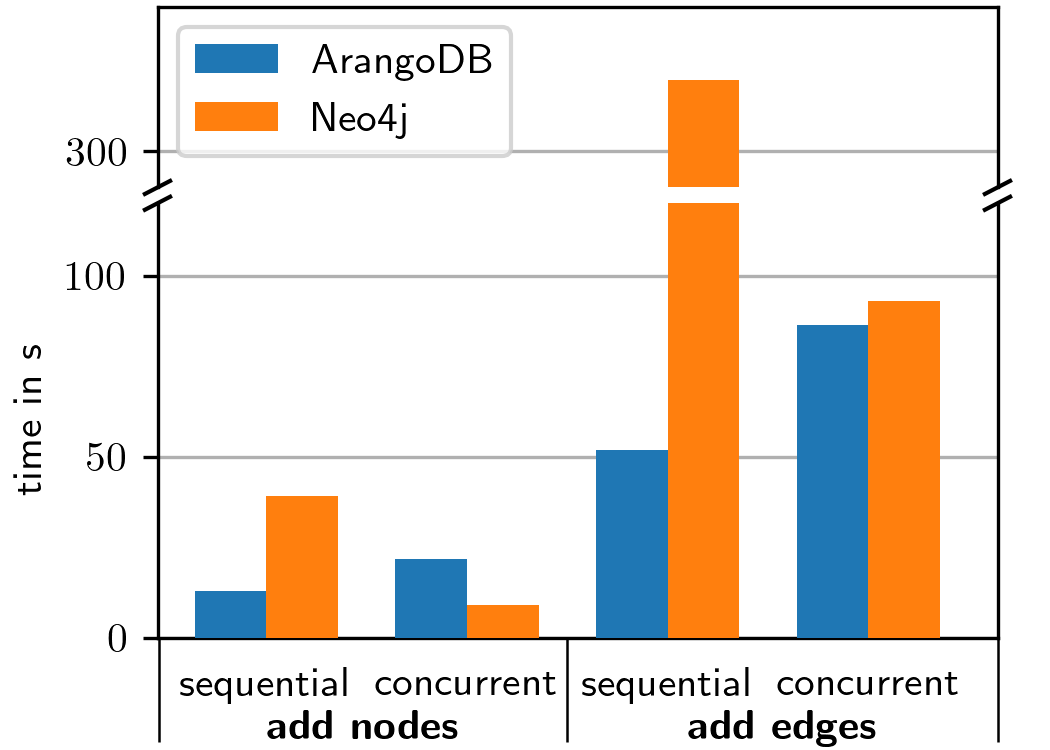}
\caption{Single import}
		\label{fig:multithreading_insert}
	\end{subfigure}
	\begin{subfigure}[b]{0.45\textwidth}
	\centering
		\includegraphics[width=\textwidth, trim=0 0.7cm 0 0.5cm,clip]{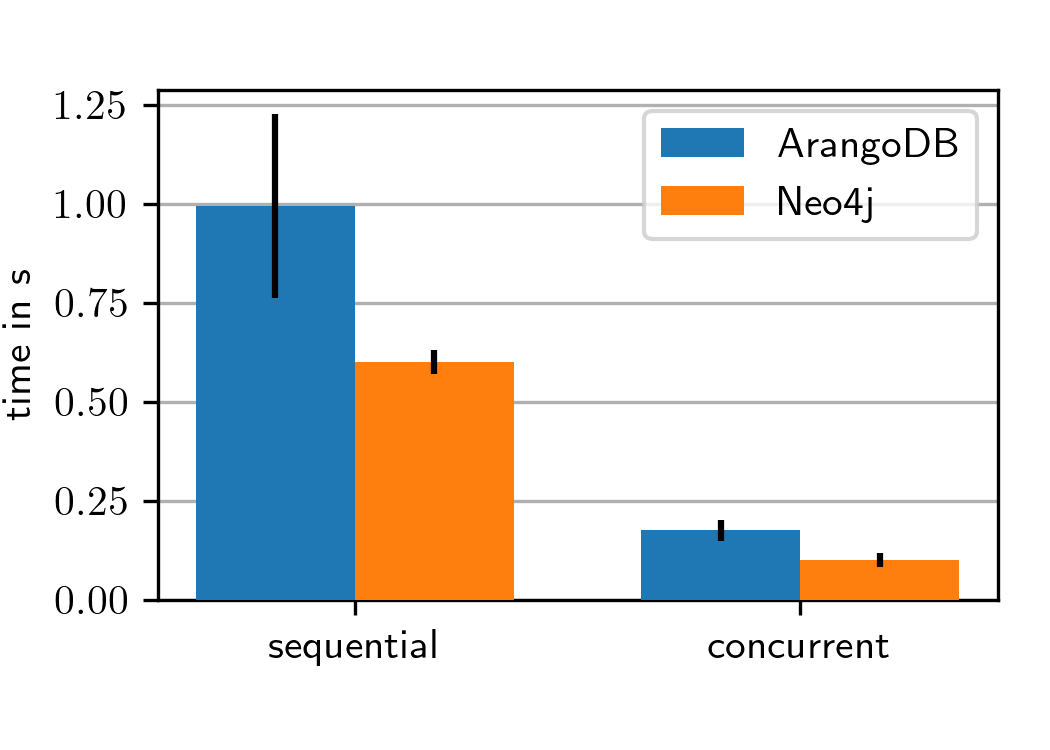}
	\caption{Path search}
	\label{fig:multithreading_search}
	\end{subfigure}
\caption{Time for importing maze data and querying a shortest path. The queries are submitted either sequentially or concurrently.}
\label{fig:multithreading}
\end{figure}
\begin{figure}[h!]
	\centering
    \includegraphics[width=0.45\textwidth, trim=0 0.2cm 0 0.7cm,clip]{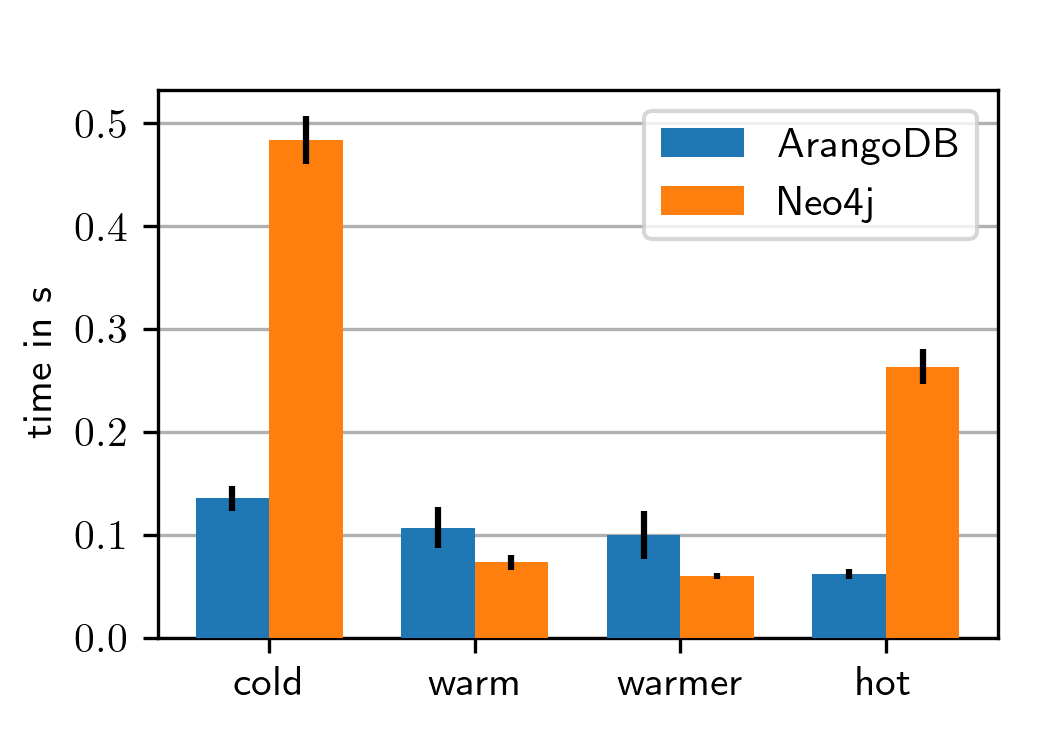}
    \caption{Influence of the cache state on the shortest path search.}
    \label{fig:cache_state}
\end{figure}
\begin{figure*}
  \includegraphics[width=0.94\textwidth, trim=1cm 1cm 0 0cm,clip]{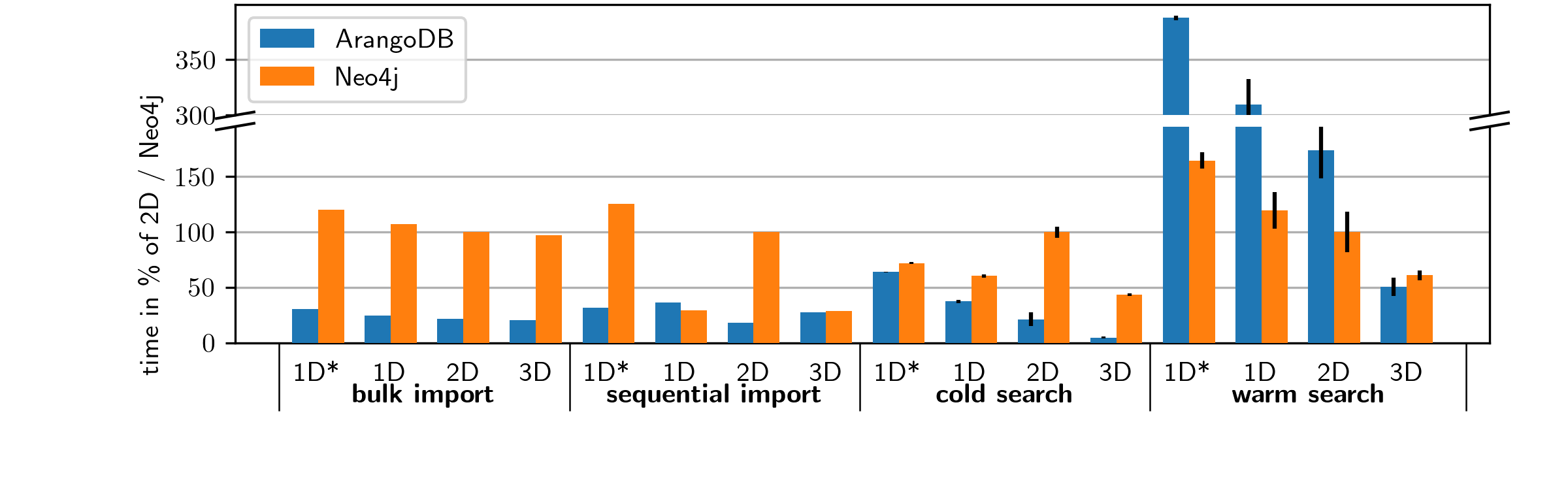}
  \caption{Influence of the maze dimensionality.}
\label{fig:dimensionality}
\end{figure*}
\FloatBarrier
\begin{figure}
	\centering
	\begin{subfigure}[b]{0.45\textwidth}
		\centering
		\includegraphics[width=\textwidth, trim=0 0.2cm 0 1.5cm,clip]{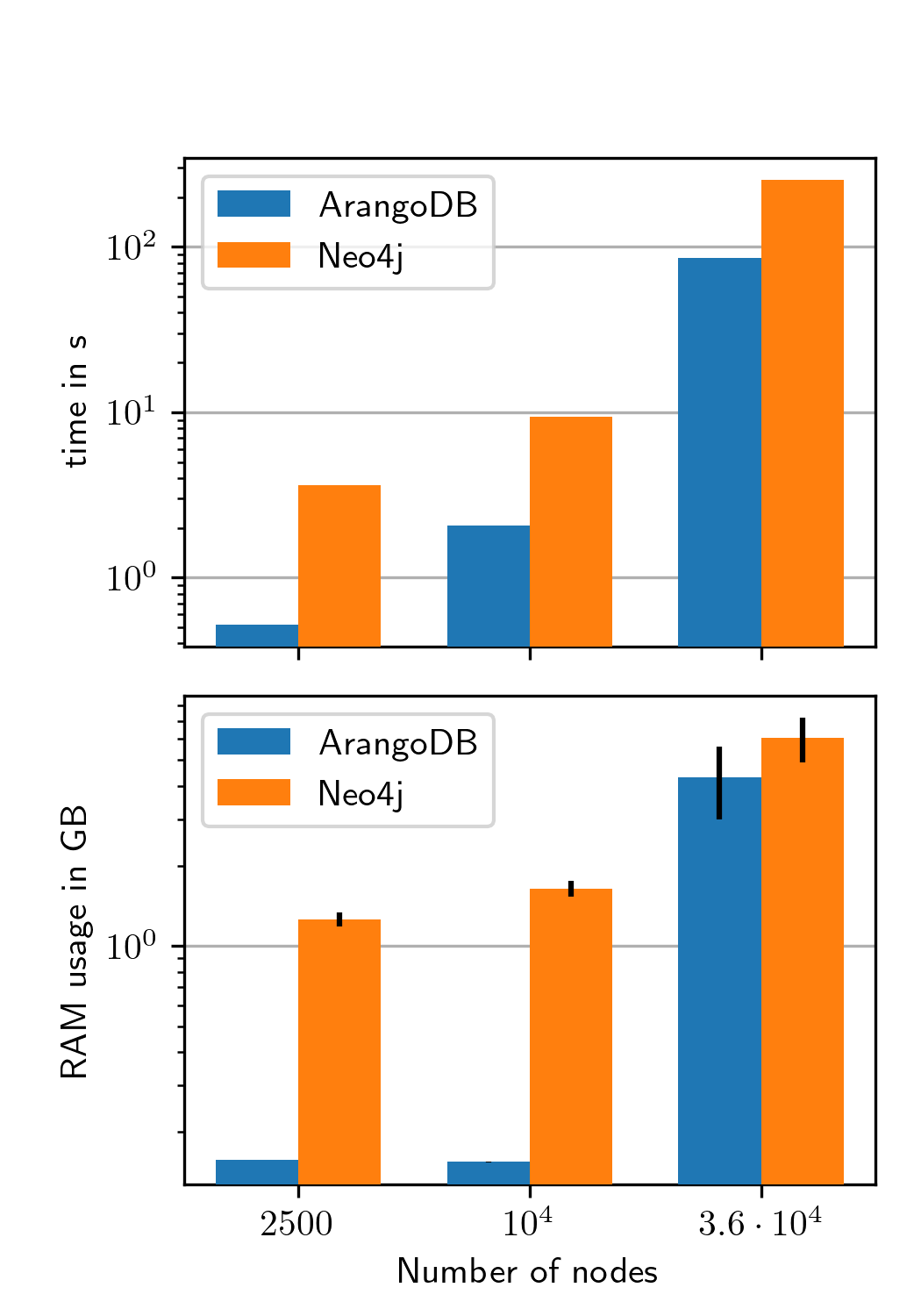}
		\caption{Bulk import}
		\label{fig:size_save}
	\end{subfigure}
	\begin{subfigure}[b]{0.45\textwidth}
	\centering
		\includegraphics[width=\textwidth, trim=0 0.2cm 0 0.5cm,clip]{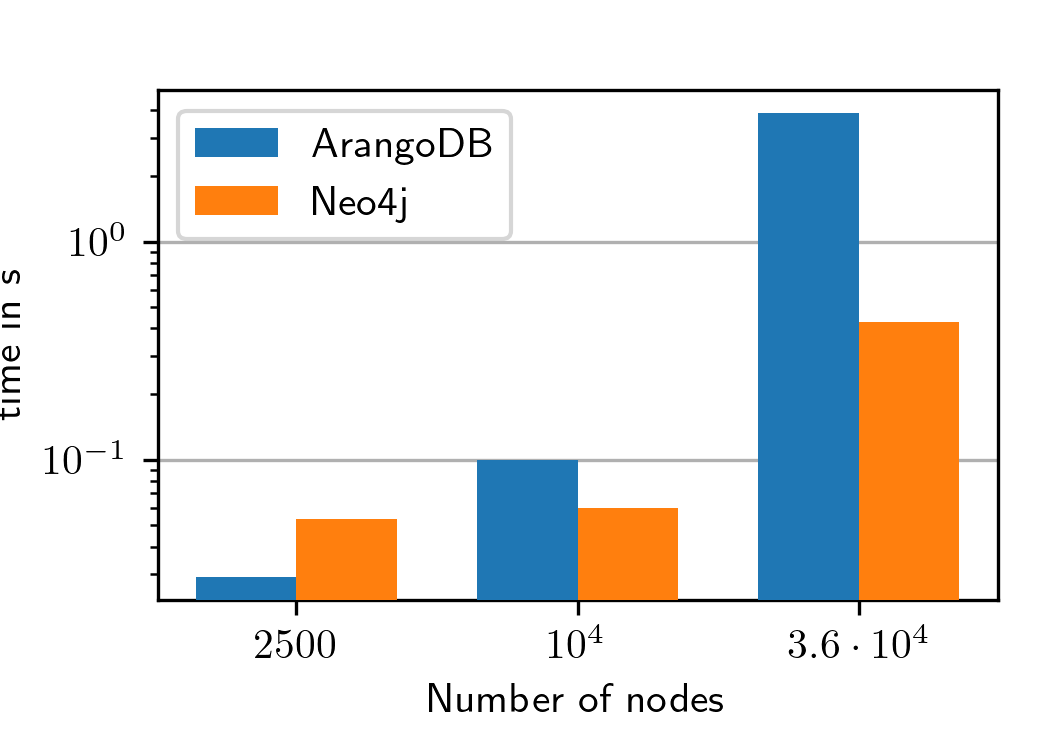}
	\caption{Path search}
	\label{fig:size_search}
	\end{subfigure}
\caption{Influence of the maze size.}
\label{fig:size}
\end{figure}	

\subsection{Concurrent processing}
The influence of concurrency is evaluated for the single import queries and for the path search. We left out the bulk import, as it is not a realistic use case to submit multiple bulk queries sequentially (rather than combining them to one bulk). 
In Fig.~\ref{fig:multithreading_insert}, the time for importing the maze data to the GDBMS is shown. With Neo4j®, both adding nodes and adding edges is significantly faster ($4.4$ and $3.35$ times respectively) when the queries are submitted concurrently. In contrast, with ArangoDB, the concurrency does not improve the performance. The computation time even increases is this case. Despite the performance improve by concurrency in Neo4j®, importing the whole data in a bulk is still more than $5$ times faster.
For the path search, shown in Fig.~\ref{fig:multithreading_search},  concurrency does decrease the processing time for both GDBMSs significantly ($5.3$ and $8.7$ times faster respectively). 

\subsection{Cache state}
As described in Sec.~\ref{sec:usage}, the cache cannot improve the performance of data import queries. Thus, only path search is evaluated.
The results are shown in Fig. \ref{fig:cache_state}. The time for finding the shortest path decreases when the cache gets warmer. While this effect is clearly visible for Neo4j®, there is no significant decrease in time for ArangoDB. 
The best performance is reached in the warm state and this performance does not improve significantly after the second query for a path search.

\subsection{Dimensionality of the maze}
To evaluate the influence of the degree of the graph, we vary the dimensionality of the maze. This way, the degree varies from $1$ edge per node (1D* with unidirectional edges), $2$ edges per node (1D) to $6$ edges per node (3D). This can have an effect on the queries for importing the edges, as the nodes are ordered differently in memory then. It obviously has an effect on the path search, as the number of search directions increases with the degree of the graph. 
The size of the mazes is chosen such that the number of edges is approximately equal to $4\cdot10^4$ for all dimensionalities. The evaluated maze sizes are:
\begin{itemize}
    \item 1D* maze with $4\cdot10^4$ cells ($4\cdot10^4$ nodes)
    \item 1D maze with $2\cdot 10 ^4$ cells ($2\cdot10^4$ nodes)
    \item 2D maze with $100\times 100$ cells ($10^4$ nodes)
    \item 3D maze with $19\times19\times19$ cells ($7\cdot 10^3$ nodes)
\end{itemize}

In Fig. \ref{fig:dimensionality}, the execution time of all evaluated queries is shown. For bulk importing the edges, the execution time increases with the number of nodes. Doubling the number of nodes increases the execution time by $74\%$ (ArangoDB) and $12\%$ (Neo4j®) respectively. In the sequential node, there is no clear trend identifiable. The execution time of a path search query, when the cache is warm, significantly decreases with increasing dimensionality. This trend is comprehensible, as the number of nodes decreases, however, it also implies that searching the trivial path in the $1$D maze takes a significant amount of time that could be eliminated by improving the data representation.
When the cache is cold, the behavior is similar in ArangoDB. Whereas in Neo4j®, surprisingly, the query takes longest for the $2$D maze. 

The time for importing the $1D*$ graph (with $4\cdot10^4$ nodes and edges) takes $10\%$ more time as importing the real road network of Amberg (with $2\cdot10^4$ nodes and $3.4\cdot 10^4$ edges). The path search takes more than $10$ times more time than the path search in the example in Amberg, which results from the larger distance between start and goal. For a distance of $100$ nodes, the search time reduces to $0.007$s in the $1D*$ array.

\subsection{Size of the maze}
To evaluate the influence of the computation time on the size of the maze, we run tests for mazes of the size $50\times50$ ($2500$ nodes), $100\times 100$ ($10^4$ nodes) and $190\times190$ ($3.6\cdot 10^4$ nodes). 
In Fig. \ref{fig:size_save}, the computation time and RAM usage of the bulk import of all nodes and edges is shown. In ArangoDB, for when number of nodes reduces to a quarter, also the computation time is divided by four. For the larger maze however, the computation time increases much more, namely by the factor of $42$. Neo4j® shows a similar behavior, although decreasing the maze size decreases the computation time even less. A similar behavior is apparent in the RAM usage. It is noteworthy a this point that the idle RAM usage (cold cache and no queries) of Neo4j® is significantly larger with $1.2$GB than the idle RAM usage of ArangoDB with $0.15$GB.

In Fig. \ref{fig:size_search}, the computation time of the path search query is shown for the different sizes of the maze. In ArangoDB, again, the computation time is divided by four for the smaller maze and increased by the factor $36$ for the larger maze. In Neo4j® the computation time decreases only by $11\%$ for the smaller maze. For the larger maze, it increases by the factor $7$. Thus, while ArangoDB is faster in the smaller mazes, Neo4j® is faster in the larger ones.



\section{Conclusion}\label{sec:conclusion}
Road networks originating from map data, as for example from OpenStreetMap, can easily be converted to graphs that can be managed by the GDBMSs evaluated in this study. Both ArangoDB and Neo4j® offer efficient algorithms for finding the shortest path in these graphs. This offers many opportunities for utilizing them for route finding in different kinds of applications. 
The GDMSs are designed for large amounts of data and also for handling queries concurrently. However, to get the best out of them and to correctly assess the performance and the scalability of the system, it is important to consider several aspects regarding characteristics of the data and the usage patterns. Our results show how the computation time changes when introducing concurrency, when warming up the cache or when the structure or size of the data changes. The results of the tests on generated data sets can be applied to real data, e.g. from road networks of OSM.


\addtolength{\textheight}{-12cm}   





\bibliography{IEEEabrv,template}

\end{document}